\documentclass[twocolumn,showpacs,superscriptaddress,preprintnumbers,amsmath,amssymb]{revtex4-1}
\setcitestyle{super}
\usepackage{graphicx}
\usepackage{dcolumn}
\usepackage{appendix}
\usepackage{bm}
\usepackage{enumitem}
\usepackage{booktabs}

\usepackage{natbib}

\usepackage[usenames,dvipsnames]{xcolor}
\usepackage{subfigure}
\usepackage{bbm}
\usepackage{enumerate}
\usepackage[
  bookmarks=true,
  colorlinks,
  linkcolor=black,
  urlcolor=black,
  citecolor=black,
  plainpages=false,
  pdfpagelabels,
  final,
  breaklinks=true
]{hyperref}

\usepackage{titlesec}
\usepackage{array}
\usepackage{dsfont}
\usepackage{physics}
\usepackage{multirow}
\usepackage{pifont}
\usepackage{soul}

\begin{document}

\title{A Fully Dynamical Description of Time-Resolved Resonant Inelastic X-ray Scattering of Pyrazine}

\author{Antonia Freibert}
\email{afreiber@physnet.uni-hamburg.de}
\affiliation{Department of Physics, University of Hamburg, Luruper Chaussee 149, 22761 Hamburg, Germany}

\author{David Mendive-Tapia}
\affiliation{Theoretical Chemistry, Institute of Physical Chemistry, Heidelberg University, Im Neuenheimer Feld 229, 69120, Heidelberg, Germany}

\author{Oriol Vendrell}
\email{oriol.vendrell@uni-heidelberg.de}
\affiliation{Theoretical Chemistry, Institute of Physical Chemistry, Heidelberg University, Im Neuenheimer Feld 229, 69120, Heidelberg, Germany}

\author{Nils Huse}
\affiliation{Department of Physics, University of Hamburg, Luruper Chaussee 149, 22761 Hamburg, Germany}

\date{\today}

\begin{abstract}
Recent advancements in ultrashort and intense X-ray sources have enabled the utilisation of resonant inelastic X-ray scattering (RIXS) as a probing technique for monitoring photoinduced dynamics in molecular systems. To account for dynamic phenomena like non-adiabatic transitions across the entire electronic state manifold, a time-dependent framework is crucial. Here, we introduce a fully time-dependent approach for calculating transient RIXS spectra using wavepacket dynamics simulations, alongside an explicit treatment of the X-ray probe pulse that surpasses Kramers-Heisenberg-Dirac constraints. Our analysis of pyrazine at the nitrogen K-edge underscores the importance of considering motion effects in all electronic states involved in the transient RIXS process. As a result, we propose a numerically exact approach to computationally support and predict cutting-edge time-resolved RIXS experiments.
\end{abstract}

\maketitle
\section{Introduction}
Resonant inelastic X-ray scattering (RIXS)\cite{Gelmukhanov1999,Ament2011} is a Raman spectroscopy technique that exploits X-ray resonances to create an intermediate core-excited state, followed by a spontaneous photon emission that propels the quantum system to a lower excited state. The energy difference between incident and emitted X-ray photon provides an excitation spectrum that is not only element-specific but also specific to the particular chemical environment of an atom. Moreover, RIXS can populate final excited states that cannot be directly probed by conventional one-photon absorption spectroscopy due to restrictions imposed by dipole selection rules.\cite{Gelmukhanov1994,Skytt1995,Luo1995,Sun2011} Moreover, in the condensed phase, where solids and solvents prohibit measurement of UV absorption spectra beyond $\lesssim$ 9~eV, RIXS allows accessing spectra up to energies in the extreme ultraviolet, i.e. inner-valence excitations.\cite{Ochmann2022} As the spectral resolution is not constrained by the core-hole lifetime broadening of the initial core-excitation, transition bands in RIXS spectra can exhibit vibrational substructure, providing information on the underlying nuclear dynamics and corresponding potential energy surfaces.\cite{Hennies2010,Pietzsch2011,VdCruz2017,soderstrom2020,Eckert2021} RIXS with hard X-rays usually targets core-excited final states because of the very low yield of valence-excited final states (resulting from so-called {\it valence-to-core} transitions of electrons that fill the core-hole). RIXS studies with soft X-rays generally target final states that are valence- or vibrationally excited (or of other nature such as magnons). The versatility of RIXS to valence-excited final states is demonstrated by its application in various studies, encompassing molecules,\cite{Marchenko2015,Zitnik2015,Lundberg2020} liquids,\cite{Blum2012,Weinhardt2013,Kjellsson2020,Young2021} and solids.\cite{Kotani2001,Revelli2019,Magnaterra2023}

Initially enabled by synchrotron radiation X-ray sources with increased spectral brightness,\cite{sync2} RIXS experienced a further advancement with the emergence of X-ray free-electron lasers (XFELs).\cite{McN10:814,Pel16:15006,Zha17:,Sed17:115901} This development facilitated the extension of RIXS to time-resolved studies,\cite{Beye2013a,Lu2020,Gel2021} exploiting nonlinear processes in optical pump/RIXS probe spectroscopy with applications in fields ranging from condensed matter physics\cite{Mitrano2020} to molecular chemistry\cite{Lundberg2020}. Within the present work, we focus on chemical applications where recent successful inquiries into photodissociation dynamics\cite{Wernet2015,Kunnus2016,Kunnus2016b}, charge-transfer excitations\cite{Jay2018,Norell2018,Jay2021}, and excited-state proton transfer\cite{Eckert2017} underscore the potential of time-resolved RIXS as a spectroscopic tool that provides element-specific Raman spectra of transient species on ultrafast timescales from vibrational frequencies to the vacuum ultraviolet in the condensed phase.

The complex nature of the underlying processes necessitates the utilisation of advanced computational techniques to decode mechanistic information from experimental data, aid in their analysis, and even forecast complicated spectral features. Consequently, a range of methods\cite{Norman2018}, both time-independent and time-dependent, have emerged to describe (steady-state) RIXS spectra for molecules where conventional time-independent approaches typically rely on the established Kramers-Heisenberg-Dirac formula employing an eigenstate representation of the system in the frequency domain.\cite{Kramers1925,Dirac1927} Conversely, time-dependent methods based on wavepacket propagation compute the scattering amplitude by the time-dependent overlap between initial and final vibronic states providing an equivalent time domain approach under steady-state conditions.\cite{Lee1979,Tannor1982} 

Previous \textit{in silico} time-resolved RIXS studies of molecules considered pump-induced states as isolated quasi-static snapshots, for which steady-state RIXS spectra were simulated.\cite{Kunnus2016,Kunnus2016b,Jay2021,Banerjee2024} This approach enabled identification of contributions from the consideres chemical species to the overall RIXS spectrum in a simplified manner, neglecting real-time motion effects. However, when a molecule is impulsively excited, the created wavepacket will evolve in time depending on the shape of the corresponding potential energy surfaces. These dynamics and interplay between the resulting population transfers can significantly impact the nuclear motion and therefore the overall experimental signal measured. Hence, for a description that incorporates nuclear motion effects arising from both the pump-induced valence and intermediate core-excited state dynamics, as well as from dynamics of the final ground and valence-excited states, a time-dependent framework is required that is able to describe non-adiabatic phenomena.

In this work, we present a full quantum dynamical treatment of femtosecond (fs-) RIXS spectroscopy employing the multiconfiguration time-dependent Hartree (MCTDH) method. For this purpose, we utilise a vibronic coupling Hamiltonian encompassing all electronic states involved in the fs-RIXS process as well as an explicit description of the coherent X-ray probe pulse. We conduct calculations of transient RIXS spectra at the nitrogen K-edge of pyrazine. Pyrazine is chosen as an interesting likewise complex benchmark system due to its pronounced non-adiabatic behavior in both the valence and core excited state manifolds. Through explicit assignment of specific spectral features to nuclear motion, we highlight the significance of adopting a comprehensive time-domain approach to achieve an accurate description of time-resolved RIXS signals including all dynamical dimensions.
\section{Methodology}
\subsection{Quantum Dynamics}
The time-dependent Schr\"odinger equation for the nuclear dynamics is solved employing the Multi-Layer (ML-) MCTDH\cite{Wang03,Manthe08,Vendrell11} wavepacket propagation method. The total MCTDH\cite{Beck00,Mey09:} wavefunction \textit{ansatz} reads
\begin{align}\label{eq:mctdhansatz}
    \Psi (\mathbf{Q}, t)  = \sum\limits_{j_1 ... j_f}A_{j_1 ... j_f}(t)\prod\limits_{i=1}^f\varphi_{j_i}(Q_i,t) 
\end{align}
where $A_{j_1 ... j_f}$ are expansion coefficients of the low-dimensional single particle functions (SPFs) $\varphi_{j_i}$. Using a linear combinations of time-independent primitive basis functions $\chi_{r_i}$, the time-dependent SPFs are themselves represented by 
\begin{align}
    \varphi_{j_i}(Q_i,t) = \sum\limits_{r_i}c_{r_i j_i}(t)\chi_{r_i}(Q_i)
\end{align}
providing a discrete variable representation (DVR) grid. The equations of motion are finally derived by applying the Dirac-Frenkel variational principle\cite{Dirac1930,Fre34:}.

Within ML-MCTDH, the SPFs are recursively expanded in the form of Eq \eqref{eq:mctdhansatz} until the last layer of time-dependent SPFs is represented by the discrete time-independent grid. This variant of MCTDH provides an efficient tool to treat larger systems fully quantum mechanically.
\subsection{Vibronic Coupling Hamiltonian}
Justified by the large energy gap, we completely decouple the valence- and core-excited states concerning non-adiabatic transitions. Moreover, the group approximation is employed separating both, the valence- and core-excited state manifold of pyrazine, into individual subsets of states, resulting in the subsequent matrix representation of the molecular Hamiltonian
\begin{align}
\mathbf{H}_{mol} =  \begin{pmatrix}
\mathbf{H}_v & 0\\
0 & \mathbf{H}_c
\end{pmatrix} 
\end{align}
where $\mathbf{H}_v = \mathrm{diag}(\mathbf{H}_{v_1},\mathbf{H}_{v_2}) $ and $\mathbf{H}_c =\mathrm{diag}(\mathbf{H}_{c_1} , \mathbf{H}_{c_2} )$ denote the sub-Hamiltonians acting on the valence- and core-excited state manifolds, respectively. In order to avoid singularities in the non-adiabatic coupling terms,\cite{Koeppel84,Worth04} we represent the molecular Hamiltonian $\mathbf{H}_{mol}$ in a diabatic electronic basis such that each sub-Hamiltonian can be written as
\begin{align}
\mathbf{H}_{x} &= \hat{T}\mathbf{1} + \mathbf{W}_x , \ x\in\{ v_1,v_2,c_1,c_2\}
\end{align}
with the kinetic energy operator $\hat{T}$ and the diabatic potential matrix $\mathbf{W_x}$. We further invoke a vibronic coupling model\cite{Cederbaum1977,Domcke1977,Koeppel84} where the diabatic potentials are expanded as Taylor series around the Franck-Condon point
\begin{align}
   \mathbf{H}_x = \mathbf{H}^{(0)} + \mathbf{W}_x^{(0)} + \mathbf{W}_x^{(1)} + \mathbf{W}_x^{(2)} + ...
\end{align}
where the zero-order Hamiltonian is the harmonic ground-state Hamiltonian
\begin{align}
\mathbf{H}^{(0)} = \sum\limits_i\frac{\omega_i}{2}\left( -\frac{\partial^2}{\partial Q_i^2} + Q_i^2\right)\mathbf{1}
\end{align}
with $\omega_i$ representing the frequency of mode $Q_i$. The diabatic potential matrix up to first order is defined by
\begin{align}
\begin{split}
    \mathbf{W}_x^{(\alpha\beta)}(\mathbf{Q}) = E^{(\alpha)}\delta_{\alpha\beta} &+ \sum\limits_i\kappa_i^{(\alpha)}Q_i\delta_{\alpha\beta} \\
   &+ \sum\limits_i \lambda_i^{(\alpha\beta)}Q_i(1-\delta_{\alpha\beta})
\end{split}
\end{align}
with the vertical excitation energies $E^{(\alpha)}$ to the $\alpha$-th electronic state and the linear intra- and interstate coupling parameters
\begin{align}
    \kappa_i^{(\alpha)} &=\left. {\frac{\partial \braket{\Psi_\alpha| \mathbf{H}_{el} |\Psi_{\alpha}}}{\partial Q_i}} \right|_{\mathbf{Q}=\mathbf{Q}_0} \\
    \lambda_i^{(\alpha\beta)} &=\left. {\frac{\partial \braket{\Psi_\alpha| \mathbf{H}_{el} |\Psi_{\beta}}}{\partial Q_i}} \right|_{\mathbf{Q}=\mathbf{Q}_0}
\end{align}
between state $|\alpha\rangle $ and $|\beta\rangle $. In order to include changes in the frequencies, we also include on-diagonal quadratic intrastate coupling constants $\gamma^{(\alpha)}$ for $\mathbf{H}_{v_1}$ and $\mathbf{H}_c$ representing the most relevant excited states in this investigation.

For strongly anharmonic modes the harmonic expression of the diabatic potentials are replaced by state-specific quartic or Morse potentials with on-diagonal matrix elements
\begin{align}
{W}^{(\alpha\alpha )}(Q_i) &=E^{(\alpha)} + \frac{1}{2}\left( \omega_i + \gamma_i^{(\alpha)} +\varepsilon_i^{(\alpha)}Q_i^2\right) Q_i^2 \\
{W}^{(\alpha\alpha )}(Q_i) &=E^{(\alpha)} + D_0^{(\alpha)}\left\{ 1 - \exp\left( -a_i^{(\alpha)}(Q_i -Q_0 )\right) \right\}^2 
\end{align}
respectively, where $\varepsilon_i^{(\alpha)}$ denotes the quartic expansion coefficient, $D_0^{(\alpha)}$ is the state-specific dissociation energy, $a_i^{(\alpha)}$ defines the curvature of the potential, and $Q_0$ is the equilibrium position.
\subsection{Pump-probe Spectra}
We use a semiclassical approach, where the full laser-driven Hamiltonian $\mathbf{H}$ reads
\begin{align}
    \mathbf{H}(t) =\mathbf{H}_\mathrm{mol} + \mathbf{H}_\mathrm{int}(t)
\end{align}
with the classical light-matter interaction operator $\mathbf{H}_\mathrm{int}$ acting as a perturbation to the molecular quantum Hamiltonian $\mathbf{H}_\mathrm{mol}$. Within the dipole and Condon approximation and assuming a single polarisation direction, the coupling to the external field is given by
\begin{align}
    H^{(\alpha\beta)}_\mathrm{int}(t) = -\mu_{\alpha\beta}\cdot\mathcal{E}(t)
\end{align}
where $\mu_{\alpha\beta}$ is the transition dipole moment. The total electric field 
\begin{align}
    \mathcal{E}(t) = \mathcal{E}_{\mathrm{pu}}(t;t_0=0) + \mathcal{E}_{\mathrm{pr}}(t;\tau ) 
\end{align}
is composed of an ultrashort $\delta$-like pump pulse $\mathcal{E}_{\mathrm{pu}}$ at time $t_0=0$ fs triggering the valence-excited state dynamics and a second pulse $\mathcal{E}_{\mathrm{pr}}$ probing the induced dynamics after a specific time delay $\Delta\tau$.

When calculating transient X-ray absorption spectra, we also assume a $\delta$-like probe pulse striking the system at time $\tau > 0$. This induces transitions from the valence- to the core-excited state manifold with Hamiltonians $\hat{H}_{v}$ and $\hat{H}_c$, respectively. The absorption cross section can then be obtained by the Fourier transform
\begin{align}\label{eq:trxas}
    I_\mathrm{XAS}(\omega ,\tau )\propto \omega\ \mathrm{Re}\int\limits_0^{\infty}\mathrm{d}t \exp{(i\omega t)}C(t;\tau)
\end{align}
of the dipole-dipole correlation function
\begin{align}
    C(t;\tau) & = \langle \Psi_v (\tau) | e^{i\hat{H}_{v} (t-\tau)}\hat{\mu}_{vc}e^{i\hat{H}_c(t-\tau)}\hat{\mu}_{cv} | \Psi_v (\tau )\rangle \\
    & = \langle \Psi_v (t) |\hat{\mu}_{vc}|\Psi_c (t)\rangle
\end{align}
where $\Psi_v$ and $\Psi_c$ denote wavepackets in the valence- and core-state manifolds, respectively, and $\hat{\mu}_{cv}$ is the transition dipole moment operator.\cite{Lee1989xas}

In absorption spectroscopy, $\delta$-pulses are conceptualised as idealisations, allowing for simultaneous interrogation of the entire spectral range. On the contrary, the RIXS process depends on narrow-band excitations to precisely define the incoming photon energy. Employing RIXS as a probe for non-stationary systems thus requires an excitation pulse that strikes a balance: it must be long enough to mitigate uncertainties in the excitation process yet short enough to achieve adequate time resolution for probing the underlying dynamics. In this work, we use Gaussian-shaped X-ray probe pulses
\begin{align}\label{eq:probe}
   \mathcal{E}_{\mathrm{pr}} (t;\tau ) &= \mathcal{A}(t-\tau) \cos\left(\omega_I(t-\tau )\right)\\
   &= \frac{1}{\sqrt{2\pi\sigma^2}} \exp\left( -\frac{(t-\tau )^2}{2\sigma^2}\right) \cos\left(\omega_I(t-\tau )\right)
\end{align}
centered at time $\tau$ with carrier frequency $\omega_I$ and temporal full width at half maximum (fwhm) duration $F_t=2\sqrt{\mathrm{ln}2}\sigma$ of the intensity profile invoking the RIXS process. Assuming no temporal overlap of the pump and probe pulse, the transient RIXS signal can be calculated using the following expression for the RIXS spectrum induced by a coherent light source\cite{Freibert2024}
\begin{align}\label{eq:trrixs}
\begin{split}
      &I_\mathrm{RIXS}(\omega_S;\omega_I ,\tau) \\
    &\propto \left( |\mathcal{U}|^2\ast \int\limits_{-\infty}^{\infty} \mathrm{d}t \exp{(-i\tilde{\omega}_S t)}\langle \tilde{\mathcal{R}}( \cdot ;\tau)|\tilde{\mathcal{R}}( \cdot ,t;\tau)\rangle\right) (\tilde\omega_I)
\end{split}
\end{align}
where $\tilde{\omega}_S = \omega_S -\tilde{\omega}_I$ is the energy loss of the system with $\tilde{\omega}_I =\omega_I + E_i$ and scattering frequency $\omega_S$. Further, $\ast$ denotes the convolution and $\mathcal{U}$ is the Fourier transform
\begin{align}
    \mathcal{U}(\omega ) = \mathcal{F}^{-1}(\mathcal{A}(t-\tau)) = \exp\left( -i\omega\tau \right)\cdot \exp\left( -\frac{\sigma^2\omega^2}{2} \right)
\end{align}
 of the Gaussian-envelope function $\mathcal{A}$ of the probe pulse defined in Eq. \eqref{eq:probe}. Furthermore, the evolving wavepacket $|\tilde{\mathcal{R}}(\omega , t; \tau)\rangle = e^{-i\hat{H}_v t}|\tilde{\mathcal{R}}(\omega ;\tau)\rangle$ is defined by the core-to-valence state projection $|\tilde{\mathcal{R}}(\omega ; \tau)\rangle =\mu_{vc} | \mathcal{R}(\omega ;\tau)\rangle$ of the Raman wavefunction
\begin{align}
    |\mathcal{R}(\omega ;\tau)\rangle = \int\limits_0^{\infty} \mathrm{d}t\  e^{-i(\hat{H}_c -i\frac{\Gamma_c}{2})t}\mu_{cv}e^{i(\tilde{\omega}_I - \omega)(t-\tau )}|\Psi_v (t)\rangle
\end{align}
collecting all dynamical information prior to the scattering event. In particular, the Raman wavefunction is a time-independent intermediate state that only parametrically depends on $\tau$.
\section{Computational Details}
The molecular Hamiltonian is divided into four energetically separated subsets of states. All parameters for the valence excited states, contained in $\mathbf{H}_{v_1}$ and $\mathbf{H}_{v_2}$, and two lowest N-1s core-excited states in $\mathbf{H}_{c_1}$ were adapted from previous studies\cite{Sala14,Freibert2024}. The values to parameterise $\mathbf{H}_{c_2}$ were obtained by fitting the diabatic potential terms from \textit{ab initio} quantum chemistry calculation using the fc-CVS-EOM-CCSD method\cite{Vid19:3117} with the Dunning correlation consistent basis set aug-cc-pVDZ\cite{Dunning1989} matching the level of theory applied in ref.~\citenum{Freibert2024}. All electronic structure calculations were carried out using the quantum chemistry software package Q-Chem\cite{qchem5}. The fitting procedure were performed using the VCHam tools, as part of the Heidelberg MCTDH package\cite{mctdh}.

Quantum dynamics propagations were run using the ML-MCTDH method as implemented in the Heidelberg MCTDH package. The layer structure, DVR, number of grid points and SPF basis size were adapted from ref.~\citenum{Freibert2024} with an increased number of 24 electronic states. The output was written such that a sufficient time resolution and frequency span for subsequent Fourier transforms is guaranteed.
\subsection{Model Hamiltonian}
\begin{figure}[!b]
\centering
\includegraphics[width=8.3cm]{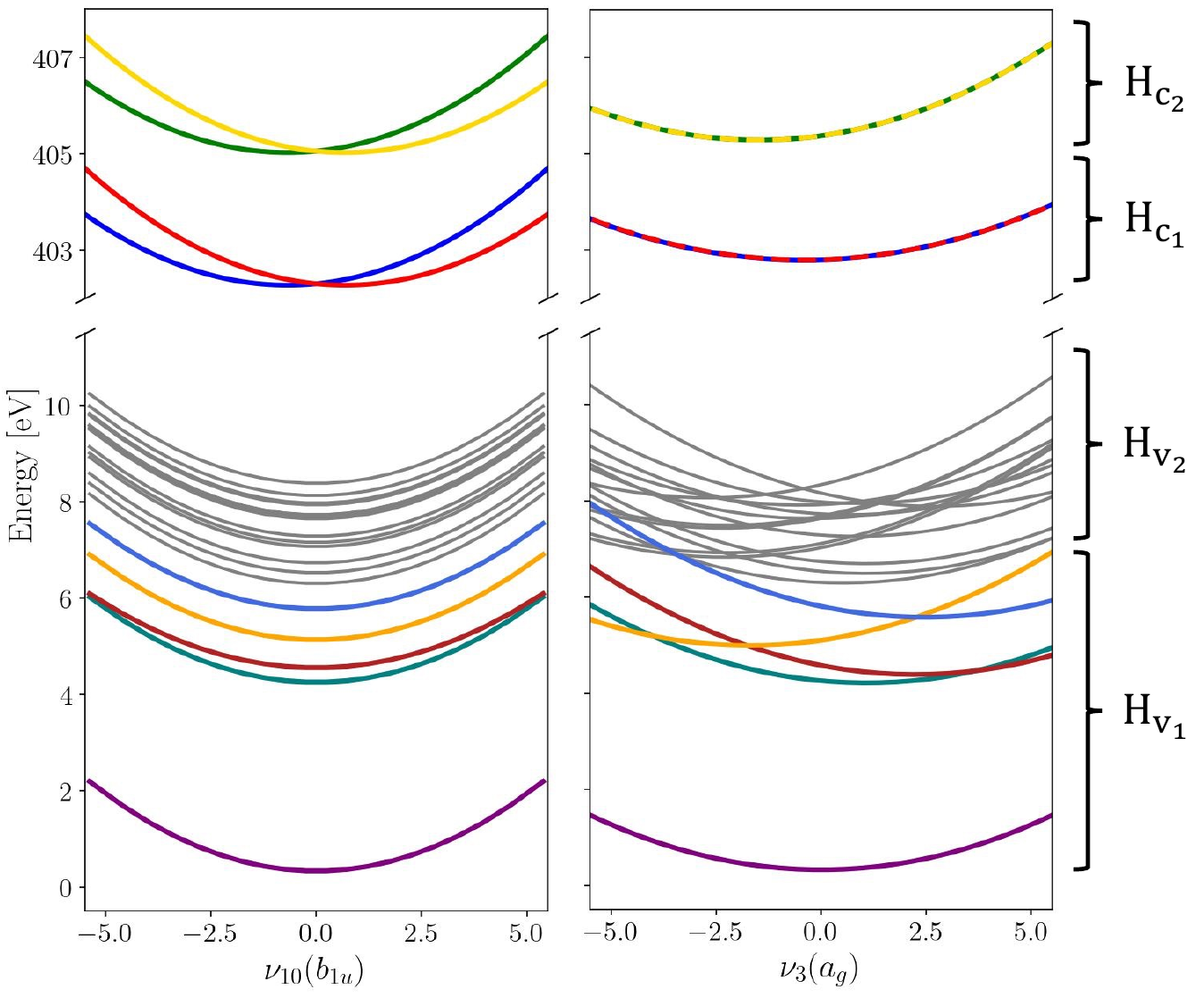}
\caption{Cuts through the diabatic potential energy surfaces along the asymmetrical $\nu_{10}(b_{1u})$ and totally symmetrical $\nu_3(a_g)$ vibrational normal modes. Electronic states that are coupled by the X-ray probe pulse are highlighted in colour while the other states that are thus playing a less significant role are held in grey for the sake of overview. Label and symmetry for each state can be found in the Supplementary Information.}\label{fig:pes}
\end{figure}
The vibronic coupling model Hamiltonian was obtained using all 24 mass- and frequency scaled normal modes of the ground state $\mathrm{D_{2h}}$ structure of pyrazine. While we utilise the (fc-CVS-)EOM-CCSD/aug-cc-pVDZ based linear vibronic coupling model for $\mathbf{H}_{v_2}$ and $\mathbf{H}_{c_1}$ from our recent study on steady-state RIXS of pyrazine\cite{Freibert2024}, we favour the vibronic model from Sala \textit{et al.}\cite{Sala14} obtained from XMCQDPT2/aug-cc-pVDZ calculations to describe $\mathbf{H}_{v_1}$ containing the ground and lowest four valence excited states. This model was constructed to simulate the UV absorption spectrum of pyrazine and yielded an excellent agreement with experimental data demonstrating its suitability to accurately describe the valence-excited state behaviour after excitation to the $B_{2u}(\pi\pi^*)$ state. In order to maintain the energy gap between $\mathbf{H}_{v_1}$ and $\mathbf{H}_{v_2}$ as given in ref.~\citenum{Freibert2024} the vertical excitation energies in $\mathbf{H}_{v_2}$ are consistently shifted by -0.6~eV. Moreover, we assume an intrinsic core-excited state lifetime of 8~fs\cite{Pri99:141} covered by an imaginary energy term in the core Hamiltonian $\mathbf{H}_c$. Figure \ref{fig:pes} shows cuts of potential energy surfaces along one of the totally symmetric normal modes and along one of the core-excited state coupling modes.

Populations in the ground and the three lowest valence-excited states give rise to strongly dipole-allowed transitions to three N-1s core-excited states where the lower two, $X_1$ and $X_2$, build $\mathbf{H}_{c_1}$. The third core excited state, $X_3$, is furthermore vibronically coupled to a nearly degenerate fourth state, $X_4$, necessitating its inclusion in the Hamiltonian $\mathbf{H}_{c_2}$. The energies and linear coupling parameters obtained in this work for $\mathbf{H}_{c_2}$ are listed in Tab.~\ref{tab:Hc2}. All transition dipole moments are given in Tab.~\ref{tab:TDM}. A complete list of parameters of the full molecular Hamiltonian can be found in ref. \citenum{Sala14} and \citenum{Freibert2024} or in the form of an operator file in the Supplementary Information to this study.
\begin{table}[ht]
\centering
\caption{Vertical excitation energies $E^{(\alpha )}$ as well as linear intra- and interstate coupling constants $\kappa_i^{(\alpha)}$ and $\lambda_{i}^{(\alpha\beta)}$, respectively, for the core-excited states contained in $\mathbf{H}_{c_2}$. All values are in eV.}\label{tab:Hc2}
\begin{tabular}{ccccc}
\hline
 & \multicolumn{2}{c}{Symmetry} & \multicolumn{2}{c}{$E$}\\
\hline\hline
$X_3$ & \multicolumn{2}{c}{$B_{1g}$} & \multicolumn{2}{c}{405.05}\\
$X_4$ & \multicolumn{2}{c}{$A_u$}    & \multicolumn{2}{c}{405.06}\\
\hline
 & $\kappa_3$ & $\kappa_{11}$ & $\kappa_{15}$ & $\kappa_{20}$ \\
\hline\hline
$X_3$ & 0.1231 & -0.0787 & -0.1236 & -0.2731  \\
$X_4$ & 0.1230 & -0.0787 & -0.1240 & -0.2754 \\
\hline
 & $\lambda_{10}$ & $\lambda_{14}$ & $\lambda_{18}$ & $\lambda_{22}$ \\
\hline\hline
($X_3 , X_4)$ & 0.0998 & 0.1087 & 0.0213 & 0.0224  \\
\hline
\end{tabular}
\end{table}
\begin{table}[ht]
\centering
\caption{Transition dipole moments $\mu_{\alpha\beta}$ between two states $|\alpha\rangle$ and $|\beta\rangle$. The transition dipole moments to valence- and core-excited states are obtained from EOM-CCSD and fc-CVS-EOM-CCSD calculations, respectively.}\label{tab:TDM}
\begin{tabular}{lc}
\hline
State transition & $\mu_{\alpha\beta}$ \\
\hline
\hline
$S_3\leftarrow S_0$ & 0.25  \\
$S_1\leftarrow S_0$ & 0.82 \\
\hline
$X_2\leftarrow S_0$ & 0.10  \\
$X_1\leftarrow S_1$ & 0.06 \\
$X_3\leftarrow S_2$ & 0.06 \\
$X_3\leftarrow S_3$ & 0.03 \\
$X_2\leftarrow S_4$ & 0.04 \\
$X_1\leftarrow S_6$ & 0.02 \\
$X_4\leftarrow S_7$ & 0.04 \\
$X_3\leftarrow S_{12}$ & 0.05 \\
$X_1\leftarrow S_{16}$ & 0.04 \\
$X_1\leftarrow S_{18}$ & 0.04 \\
\hline
\end{tabular}
\end{table}
\section{Results}
\subsection{Time-Resolved X-ray Absorption Spectra}
\begin{figure}[h]
\centering
\includegraphics[width=9cm]{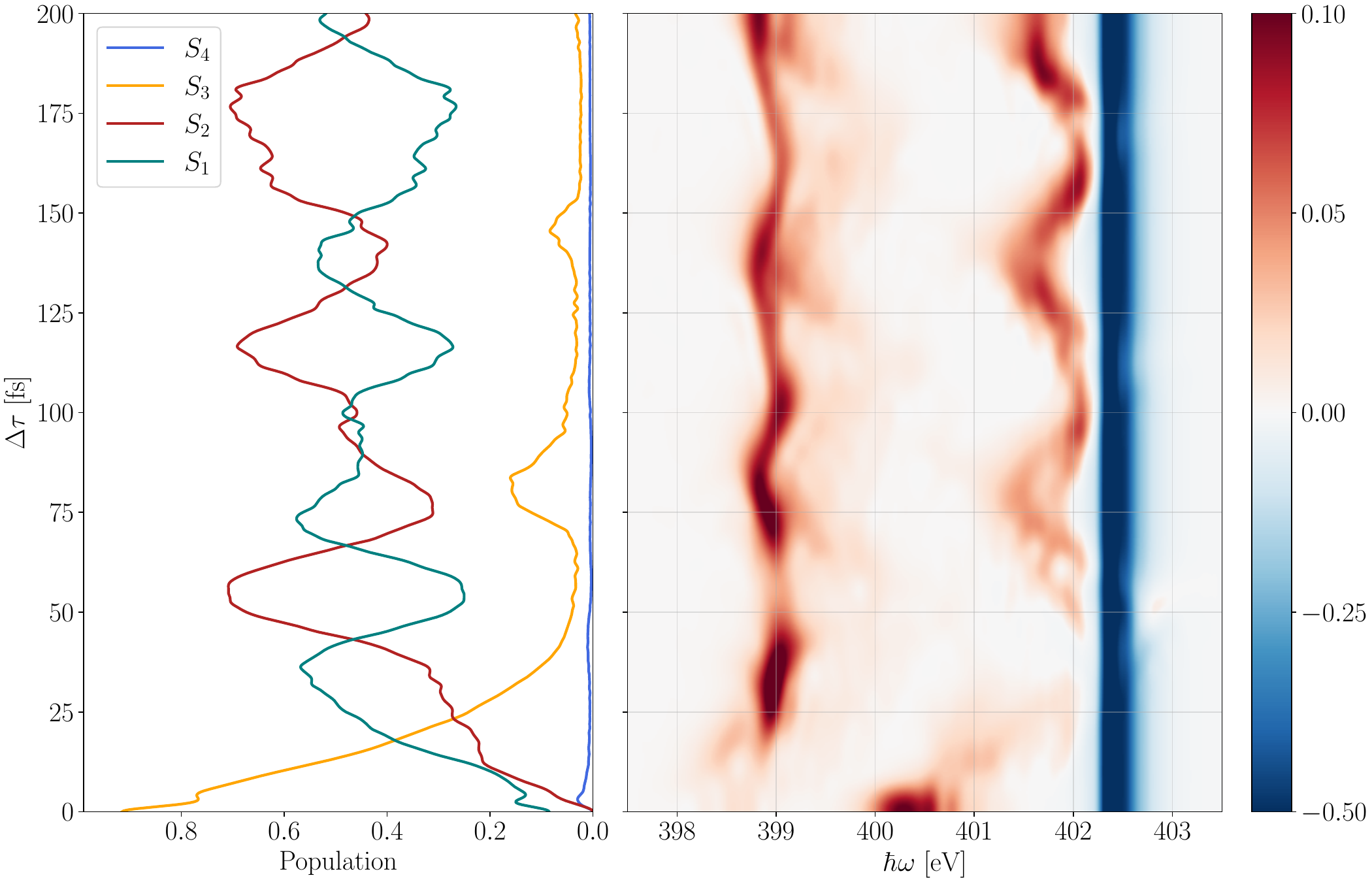}
\caption{Valence-excited state population after instantaneous, vertical excitation to the bright $S_3$ and $S_1$ states at the Franck-Condon point (left) and a three-dimensional map showing all computed differential X-ray absorption spectra for time delays between 0 and 200 fs (right).}\label{fig:trxas}
\end{figure}

To ascertain appropriate excitation energies for investigating the UV-induced dynamics of pyrazine with RIXS at the nitrogen K-edge, we compute time-resolved spectra in the X-ray absorption near-edge structure (XANES) region. For this purpose, we assume $\delta$-pulses for both the pump and probe steps and calculate the spectra according to Eq. \eqref{eq:trxas}. The diabatic valence-excited state population dynamics subsequent to the pump pulse along with a three-dimensional collection of differential absorption spectra are shown in Fig.~\ref{fig:trxas}. At time $t=0~\mathrm{fs}$, the system is primarily excited to $S_3$ with minor contributions to $S_1$ according to their transition dipole moments given in Tab. \ref{tab:TDM}. $S_3$ rapidly depopulates to both the bright $S_1$ and the dark $S_2$ state followed by oscillatory population dynamics between $S_1$ and $S_2$. This diabatic population behaviour can be traced back to low-lying conical intersections between all state pairs, $S_1/S_2$, $S_1/S_3$ and $S_2/S_3$.\cite{Sala14} 

The population dynamics are well reflected in the transient absorption spectra, where the strong negative bleach signal located around 402.5~eV corresponds to $X_2\leftarrow S_0$ transitions while the positive excitation bands at approximately 399.0~eV and 401.8~eV stem from excited state absorption. Within the first few femtoseconds, $S_3$ is mainly populated leading to an absorption band around 400.5~eV. Due to vibrational relaxation in $S_3$ as well as X-ray excitation to potentially higher vibronic states in $X_3$, the $X_3\leftarrow S_3$ transitions are then shifted to higher frequencies by about 1.3~eV where they energetically overlap with the $X_3\leftarrow S_2$ transition band. The absorption band at 399.0~eV only stems from $X_1\leftarrow S_1$ transitions. The non-adiabatic oscillatory population dynamics between $S_1$ and $S_2$ are accurately mapped in the intensity variations of the two main excited state absorption bands in good agreement with previous simulations.\cite{Freibert2021} However, it is worth noting that we formerly used a reduced pyrazine model that comprised only the 9 most dominant vibrational normal modes. In this study, a full 24-dimensional model is used leading to more delocalised wavepackets on the valence excited state manifold that manifest in broader and somewhat asymmetrical line shapes of the excitation bands. This behaviour is particularly pronounced in the strongly fluctuating absorption band around 401.8~eV which also overlaps with the bleach signal at some time delays. In contrast, our previous calculations showed this absorption band to be well separated from the bleach signal for the entire propagation time.

Based on the time-dependent differential absorption map in Fig. \ref{fig:trxas}, we choose excitation energies of 399.0~eV and 401.5~eV for the following fs-RIXS simulations, probing the dynamics on $S_1$ and $S_3/S_2$, respectively. The latter X-ray excitation energy was chosen slightly below the center of the fluctuating absorption band to exclude contributions from ground-state bleaching. Moreover, the Supplementary Information also contains fs-RIXS calculations for an excitation energy of 402.5~eV to show the behaviour in the overlap region of the bleach signal and the close-lying excited state absorption.

\subsection{Femtosecond Resonant Inelastic X-ray Scattering}
The dynamics triggered by the optical pump pulse primarily involve three valence excited states and are mainly driven by the interstate coupling parameters between the state pairs $S_1/S_3$ and $S_1/S_2$ with oscillation periods of 19~fs and 21~fs, respectively.\cite{Sala14} We therefore choose a Gaussian X-ray probe pulse with a temporal FWHM duration of $F_t \approx 8$~fs enabling to follow this ultrafast dynamics without loosing relevant information due to spectral broadening caused by a broadband excitation pulses.\cite{Freibert2024} 

\begin{figure*}
\centering
\includegraphics[width=17.1cm]{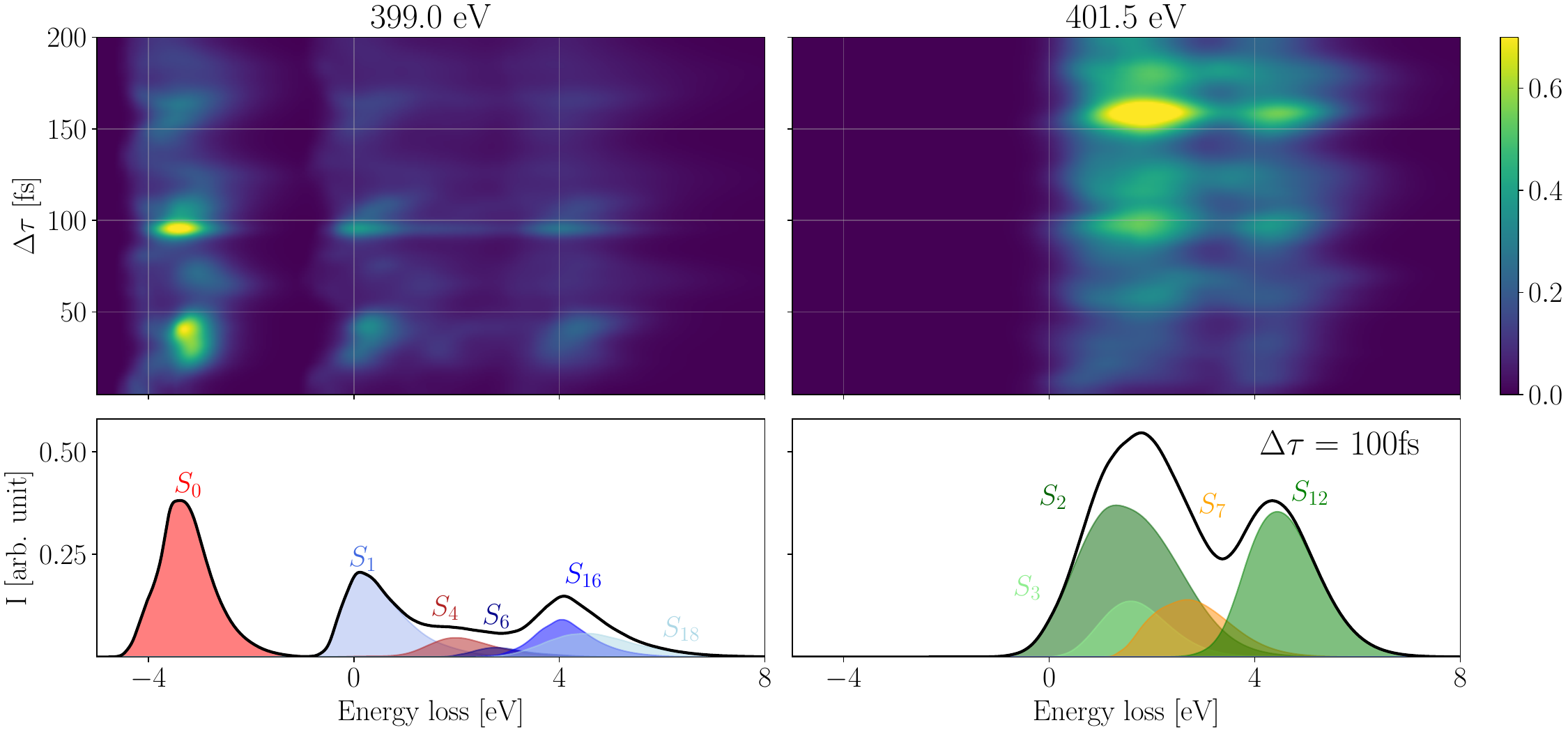}
\caption{Top: Three-dimensional collection of calculated fs-RIXS spectra for time delays $\Delta\tau$ between 5 fs and 200 fs with a step size of 5 fs using excitation energies of 399.0 eV (left) and 401.5 eV (right). Bottom: fs-RIXS spectrum at time delay $\Delta\tau = 100$ fs. Transitions stemming from $X_1$, $X_2$, $X_3$ and $X_4$ are highlighted in red, blue, green and yellow, respectively. The final electronic state of each transition is annotated in the plot.}\label{fig:fsrixs_399_401}
\end{figure*}

The transient RIXS spectra are computed using Eq.~\eqref{eq:trrixs} with carrier frequencies of 399.0~eV and 401.5~eV and at time-delays $\Delta\tau$ ranging from 5 to 200~fs. A three-dimensional collection of all spectra along with a single fs-RIXS spectrum for $\Delta\tau =100~$fs is shown in Fig.~\ref{fig:fsrixs_399_401}. Moreover, a two dimensional representation of the time evolution of the fs-RIXS spectrum in 10~fs steps for both excitation energies can be found in the Supplementary Information.

As discussed in the previous section, RIXS probing at 399.0~eV only promotes $X_1\leftarrow S_1$ transitions entailing four dipole allowed scattering transitions back to valence excited states $S_1$, $S_6$, $S_{16}$ and $S_{18}$ (see Tab.~\ref{tab:TDM}) with transition bands located at approximately 0.0~eV, 3.0~eV, 4.0~eV and 4.5~eV, respectively. These values correspond to the electronic states' energy gap to $S_1$. Moreover, the RIXS spectrum exhibits a strong anti-Stokes signal at -3.8~eV and a weaker transition band at 2.0~eV belonging to transitions to $S_0$ and $S_4$, respectively, which are dipole forbidden from the initially excited $X_1$ state. However, due to the delocalised nature of the N-1s core orbitals over the equivalent nitrogen atoms, pyrazine naturally posses nearly degenerate core-excited states, $X_1$ and $X_2$, that are vibronically coupled through asymmetrical normal modes of symmetry $b_{1u}$.\cite{Freibert2024} This coupling leads to ultrafast symmetry breaking of the system within the core-hole lifetime accompanied by a localisation of the core orbitals that finally permits the dipole allowed $S_0\leftarrow X_2$ and $S_4\leftarrow X_2$ transitions. Furthermore, besides the rather weak and spectrally overlapping signals at higher energy loss, the clearly distinguishable $S_0\leftarrow X_2$ and $S_1\leftarrow X_1$ transition bands can be seen over the entire time range as demonstrated in the upper left panel of Fig.~\ref{fig:fsrixs_399_401}. Consequently, after photo-excitation to $S_3$ the system leaves the immediate Franck-Condon region but remains in a configurations where the two nitrogen atoms are equivalent holding the system close to the conical intersection seam formed along the four normal modes with symmetry $b_{1u}$. These modes are responsible for disturbing the equivalence of the nitrogen atoms. This result particularly confirms the assumption that none of these vibrational normal modes plays a prominent role in the UV induced dynamics of pyrazine. They can thus be neglected in linear UV absorption simulations.\cite{Sala14} Lastly, we note that the overall signal strengths vary in accordance with the population dynamics of $S_1$ shown in Fig.~\ref{fig:trxas}.

\begin{figure*}
\centering
\includegraphics[width=17.1cm]{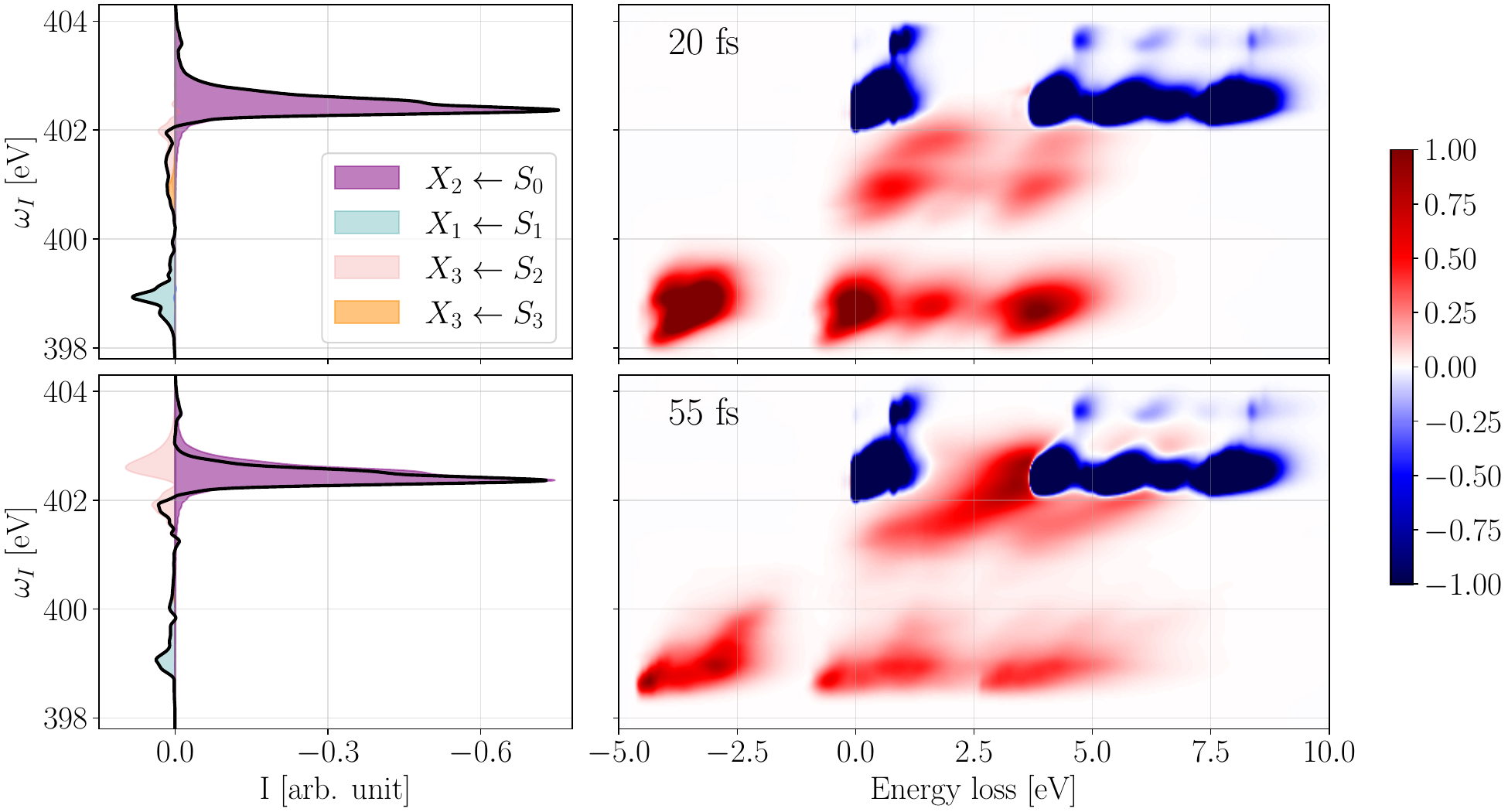}
\caption{fs-XANES (left) and corresponding fs-RIXS spectra (right) for time delays $\Delta\tau~=~20$~fs (top) and $\Delta\tau~=~55$ fs (bottom). fs-RIXS spectra are computed for excitation energies $\omega_I$ ranging from 397.0 eV to 404.9 eV with 0.1 eV stepsize.}\label{fig:fsrixs_20_55}
\end{figure*}
X-ray excitations at 401.5~eV addresses $X_3\leftarrow S_2$ as well as $X_3\leftarrow S_3$ transitions where the latter only significantly contribute within the first 20-30~fs due to the rapid depopulation of $S_3$. The related fs-RIXS spectra show two dominant spectral bands where the lower band actually originates from three energetically overlapping transitions. As before, there are two nearly degenerate core-excited states, $X_3$ and $X_4$, that are vibronically coupled along the $b_{1u}$ normal modes with coupling strengths listed in Tab.~\ref{tab:Hc2}. Due to their large overlap, ultrafast nuclear motion cannot be separately attributed to individual transitions within this spectral region. Contrary to probe excitations at 399.0~eV, the transient RIXS spectra at 401.5~eV do not contain anti-Stokes signals. Thus, the absence of transitions to the ground state indicate the symmetry of the core-excited state at 401.5~eV. Moreover, an increase of the overall signal strength can be observed for time delays $\Delta\tau$ beyond 100~fs. On the one hand, this trend can be explained by the considerable lower transition dipole moment $\mu_{X_3\leftarrow S_3}$ compared to $\mu_{X_3\leftarrow S_2}$ yielding less likely transitions at very early times, and, on the other hand by the strong modulation of the absorption band (see Fig.~\ref{fig:trxas}) resulting in a large detuning effect for time delays between 20~fs and 80~fs.

Fig.~\ref{fig:fsrixs_20_55} presents differential XANES spectra and RIXS maps as a function of excitation energy $\omega_I$ of the X-ray probe pulse and energy loss $\tilde{\omega}_S$ to the molecule at specific time delays. Both the transient absorption and the RIXS spectra exhibits positive features stemming from valence-excited state transitions and negative bleach signals related to ground-state transitions of the non-excited system. The bleach signal remains the same for all time delays as there is no radiationless relaxation mechanism from excited states back to the ground state on the femtosecond timescale. Moreover, X-ray transitions involving the ground state $S_0$ are considerably more likely compared to valence-excited state transition (see Tab.~\ref{tab:TDM}) resulting in a very dominant bleach signal at approximately 402.5~eV. Overall the shape of the excited-state transition bands is rather broad and asymmetrical for both probing techniques which must be therefore traced back to the evolution of the nuclear wavepacket of the pumped system. Additionally, the fs-RIXS maps also contain dynamical features from the intermediate core-excited states as well as from the final ground and valence-excited states.

For both time delays shown in Fig.~\ref{fig:fsrixs_20_55} the RIXS bleach signal consists of five transition bands (stemming from six core-to-valence state transitions). The higher four bands are energetically separated from the lowest transition band by about 3.9~eV, 5.3~eV, 6.5~eV and 7.5~eV. As the lowest transition band of the bleach signal is located just above 0~eV it can be uniquely assigned to the electronic ground state. Moreover, for 20 fs time delay the energy-level separation pattern of the bleach signal repeats in the excited-state signals at 399.0~eV shifted by -3.9~eV. This suggests that the corresponding initial ground and valence-excited states address the same core-excited state pair leading in turn to similar emission bands. However, the progressive valence-excited state dynamics after 55 fs smear out the recurrences, thereby making a direct assignment without additional analysis support difficult. 

The spectral proximity of the RIXS band around 401.5~eV masks parts of the time-dependent signal. Particularly at 55 fs the dominant time-independent bleach signal overlaps with the stronger part of the excited-state signal. Within the first 20 fs after valence excitation the two signals do not overlap appreciably before population transfer to $S_2$ sweeps the resonance - and hence, the RIXS signal - to higher energy. As already discussed above, the missing energy-gain features at 401.5~eV clearly indicate X-ray transitions to another pair of core-excited states.

In essence, time-resolved RIXS emerges as a dependable tool to unravel the ultrafast photophysical actions of pyrazine, augmenting the capabilities of other ultrafast X-ray spectroscopic methodos such as fs-XANES spectroscopy. Although  fs-XANES adeptly probes the dynamics induced by UV radiation, time-resolved RIXS delves deeper into dynamic facets, unveiling symmetry distortions resulting from non-adiabatic transitions within core-excited states. Furthermore, it furnishes insights into higher-lying valence excited states, which remain inaccessible in transient absorption spectroscopy due symmetry constraints or weak transition dipole moments.
\section{Conclusion}
While steady-state soft X-ray RIXS has already proven appropriate in probing the chemical and physical properties of molecular systems, time-resolved RIXS experiments have been less common due to the even more demanding spectral brightness requirements. However, advances in XFEL technology, accompanied by new instrumentation such as the hRIXS instrument at European XFEL and the ChemRIXS instrument at LCLS-II,\cite{LCLSII} hold promise of making time-resolved RIXS a routinely employed technique in the coming years. Moreover, advances in seeding technology will mean a further leap in spectral brightness of XFEL sources. To effectively extract the rich information contained within the experimental data, theoretical progress in truly describing dynamics are imperative to support these expanded experimental capabilities.

In this study, we presented a fully time-dependent quantum mechanical approach to simulating time-resolved RIXS experiments. Leveraging the capabilities of the MCTDH method, we establish a robust framework to accurately describe the intricate photophysics of pyrazine, encompassing all relevant dynamical dimensions. The interplay between population transfers within the valence-excited state manifold and the symmetry-breaking nuclear motions after X-ray interrogation can only be faithfully captured through nuclear quantum dynamics simulations that incorporate non-adiabatic phenomena. Our study further discloses challenges in building reduced-dimensionality models in larger molecules and elucidates the necessity of incorporating symmetry-breaking degrees of freedom. In particular, our time-dependent approach enables simulations beyond steady-state conditions contributing to the growing field of ultrafast RIXS experiments at pulsed X-ray sources of high spectral brightness.  
\section*{Conflicts of interest}
There are no conflicts to declare.

\section*{Acknowledgements}
A.F. acknowledges financial support from the International Max Planck Graduate School for Ultrafast imaging $\&$ Structural Dynamics (IMPRS-UFAST) and from the Christiane-N\"usslein-Vollhard-Foundation. O.V. acknowledges financial support from the German Science Foundation (DFG) through the project number 493826649. This work is supported by the Cluster of Excellence 'CUI: Advanced Imaging of Matter' of the Deutsche Forschungsgemeinschaft (DFG) - EXC 2056 - project ID 390715994 (A.F. $\&$ N.H.).

\bibliographystyle{unsrt}
\bibliography{references}{}

\begin{thebibliography}{10}

\bibitem{Gelmukhanov1999}
Faris Gel'mukhanov and Hans Ågren.
\newblock Resonant x-ray raman scattering.
\newblock {\em Physics Reports}, 312(3):87 -- 330, 1999.

\bibitem{Ament2011}
Luuk J.~P. Ament, Michel van Veenendaal, Thomas~P. Devereaux, John~P. Hill, and
  Jeroen van~den Brink.
\newblock Resonant inelastic x-ray scattering studies of elementary
  excitations.
\newblock {\em Rev. Mod. Phys.}, 83:705--767, Jun 2011.

\bibitem{Gelmukhanov1994}
Faris Gel'mukhanov and Hans \AA{}gren.
\newblock Resonant inelastic x-ray scattering with symmetry-selective
  excitation.
\newblock {\em Phys. Rev. A}, 49:4378--4389, Jun 1994.

\bibitem{Skytt1995}
Per Skytt, Jinghua Guo, Nial Wassdahl, Joseph Nordgren, Yi~Luo, and Hans
  \AA{}gren.
\newblock Probing symmetry breaking upon core excitation with resonant x-ray
  fluorescence.
\newblock {\em Phys. Rev. A}, 52:3572--3576, Nov 1995.

\bibitem{Luo1995}
Yi~Luo, Hans \AA{}gren, Faris Gel'mukhanov, Jinghua Guo, Per Skytt, Nial
  Wassdahl, and Joseph Nordgren.
\newblock Symmetry-selective resonant inelastic x-ray scattering of
  ${\mathrm{c}}_{60}$.
\newblock {\em Phys. Rev. B}, 52:14479--14496, Nov 1995.

\bibitem{Sun2011}
Y-P Sun, A~Pietzsch, F~Hennies, Z~Rinkevicius, H~O Karlsson, T~Schmitt, V~N
  Strocov, J~Andersson, B~Kennedy, J~Schlappa, A~F{\"o}hlisch, F~Gel'mukhanov,
  and J-E Rubensson.
\newblock Internal symmetry and selection rules in resonant inelastic soft
  x-ray scattering.
\newblock {\em Journal of Physics B: Atomic, Molecular and Optical Physics},
  44(16):161002, jul 2011.

\bibitem{Ochmann2022}
Miguel Ochmann, Vin{\'\i}cius Vaz~da Cruz, Sebastian Eckert, Nils Huse, and
  Alexander F{\"o}hlisch.
\newblock R-group stabilization in methylated formamides observed by resonant
  inelastic x-ray scattering.
\newblock {\em Chem. Commun.}, 58:8834--8837, 2022.

\bibitem{Hennies2010}
Franz Hennies, Annette Pietzsch, Martin Berglund, Alexander F\"ohlisch,
  Thorsten Schmitt, Vladimir Strocov, Hans~O. Karlsson, Joakim Andersson, and
  Jan-Erik Rubensson.
\newblock Resonant inelastic scattering spectra of free molecules with
  vibrational resolution.
\newblock {\em Phys. Rev. Lett.}, 104:193002, May 2010.

\bibitem{Pietzsch2011}
A.~Pietzsch, Y.-P. Sun, F.~Hennies, Z.~Rinkevicius, H.~O. Karlsson, T.~Schmitt,
  V.~N. Strocov, J.~Andersson, B.~Kennedy, J.~Schlappa, A.~F\"ohlisch, J.-E.
  Rubensson, and F.~Gel'mukhanov.
\newblock Spatial quantum beats in vibrational resonant inelastic soft x-ray
  scattering at dissociating states in oxygen.
\newblock {\em Phys. Rev. Lett.}, 106:153004, Apr 2011.

\bibitem{VdCruz2017}
Vin{\'\i}cius Vaz~da Cruz, Emelie Ertan, Rafael~C. Couto, Sebastian Eckert,
  Mattis Fondell, Marcus Dantz, Brian Kennedy, Thorsten Schmitt, Annette
  Pietzsch, Freddy~F. Guimar{\~a}es, Hans {\AA}gren, Faris Gel{'}mukhanov,
  Michael Odelius, Alexander F{\"o}hlisch, and Victor Kimberg.
\newblock A study of the water molecule using frequency control over nuclear
  dynamics in resonant x-ray scattering.
\newblock {\em Phys. Chem. Chem. Phys.}, 19:19573--19589, 2017.

\bibitem{soderstrom2020}
Johan S\"oderstr\"om, Robert Stefanuik, Franz Hennies, Thorsten Schmitt,
  Vladimir~N. Strocov, Joakim Andersson, Brian Kennedy, Justine Schlappa,
  Alexander F\"ohlisch, Annette Pietzsch, and Jan-Erik Rubensson.
\newblock Resonant inelastic x-ray scattering on ${\mathrm{co}}_{2}$: Parity
  conservation in inversion-symmetric polyatomics.
\newblock {\em Phys. Rev. A}, 101:062501, Jun 2020.

\bibitem{Eckert2021}
Sebastian Eckert, Vin{\'\i}cius Vaz~da Cruz, Miguel Ochmann, Inga von Ahnen,
  Alexander F{\"o}hlisch, and Nils Huse.
\newblock Breaking the symmetry of pyrimidine: Solvent effects and core-excited
  state dynamics.
\newblock {\em The Journal of Physical Chemistry Letters}, 12(35):8637--8643,
  2021.
\newblock PMID: 34472857.

\bibitem{Marchenko2015}
T~Marchenko, S~Carniato, L~Journel, R~Guillemin, E~Kawerk, M~{\v Z}itnik,
  M~Kav{\v c}i{\v c}, K~Bu{\v c}ar, R~Bohinc, M~Petric, V~Vaz da~Cruz,
  F~Gel'mukhanov, and M~Simon.
\newblock Electron dynamics in the core-excited cs2 molecule revealed through
  resonant inelastic x-ray scattering spectroscopy.
\newblock {\em Journal of Physics: Conference Series}, 635(11):112012, aug
  2015.

\bibitem{Zitnik2015}
M.~{\v Z}itnik, M.~Kav{\v c}i{\v c}, R.~Bohinc, K.~Bu{\v c}ar, A.~Miheli{\v c},
  W.~Cao, R.~Guillemin, L.~Journel, T.~Marchenko, S.~Carniato, E.~Kawerk, M.N.
  Piancastelli, and M.~Simon.
\newblock Resonant inelastic x-ray spectroscopy of atoms and simple molecules:
  Satellite features and dependence on energy detuning and photon polarization.
\newblock {\em Journal of Electron Spectroscopy and Related Phenomena},
  204:356--364, 2015.
\newblock Gas phase spectroscopic and dynamical studies at Free-Electron Lasers
  and other short wavelength sources.

\bibitem{Lundberg2020}
Marcus Lundberg and Philippe Wernet.
\newblock {\em Resonant Inelastic X-ray Scattering (RIXS) Studies in Chemistry:
  Present and Future}, pages 2315--2366.
\newblock Springer International Publishing, Cham, 2020.

\bibitem{Blum2012}
M.~Blum, M.~Odelius, L.~Weinhardt, S.~Pookpanratana, M.~B{\"a}r, Y.~Zhang,
  O.~Fuchs, W.~Yang, E.~Umbach, and C.~Heske.
\newblock Ultrafast proton dynamics in aqueous amino acid solutions studied by
  resonant inelastic soft x-ray scattering.
\newblock {\em The Journal of Physical Chemistry B}, 116(46):13757--13764, 11
  2012.

\bibitem{Weinhardt2013}
L.~Weinhardt, M.~Blum, O.~Fuchs, A.~Benkert, F.~Meyer, M.~B{\"a}r, J.D.
  Denlinger, W.~Yang, F.~Reinert, and C.~Heske.
\newblock Rixs investigations of liquids, solutions, and liquid/solid
  interfaces.
\newblock {\em Journal of Electron Spectroscopy and Related Phenomena},
  188:111--120, 2013.
\newblock Progress in Resonant Inelastic X-Ray Scattering.

\bibitem{Kjellsson2020}
L.~Kjellsson, K.~D. Nanda, J.-E. Rubensson, G.~Doumy, S.~H. Southworth, P.~J.
  Ho, A.~M. March, A.~Al~Haddad, Y.~Kumagai, M.-F. Tu, R.~D. Schaller,
  T.~Debnath, M.~S. Bin Mohd~Yusof, C.~Arnold, W.~F. Schlotter, S.~Moeller,
  G.~Coslovich, J.~D. Koralek, M.~P. Minitti, M.~L. Vidal, M.~Simon, R.~Santra,
  Z.-H. Loh, S.~Coriani, A.~I. Krylov, and L.~Young.
\newblock Resonant inelastic x-ray scattering reveals hidden local transitions
  of the aqueous oh radical.
\newblock {\em Phys. Rev. Lett.}, 124:236001, Jun 2020.

\bibitem{Young2021}
Linda Young, Emily~T. Nienhuis, Dimitris Koulentianos, Gilles Doumy, Anne~Marie
  March, Stephen~H. Southworth, Sue~B. Clark, Thomas~M. Orlando, Jay~A.
  LaVerne, and Carolyn~I. Pearce.
\newblock Photon-in/photon-out x-ray free-electron laser studies of radiolysis.
\newblock {\em Applied Sciences}, 11(2), 2021.

\bibitem{Kotani2001}
Akio Kotani and Shik Shin.
\newblock Resonant inelastic x-ray scattering spectra for electrons in solids.
\newblock {\em Rev. Mod. Phys.}, 73:203--246, Feb 2001.

\bibitem{Revelli2019}
A.~Revelli, M.~Moretti Sala, G.~Monaco, P.~Becker, L.~Bohat{\'y}, M.~Hermanns,
  T.~C. Koethe, T.~Fr{\"o}hlich, P.~Warzanowski, T.~Lorenz, S.~V. Streltsov,
  P.~H.~M. van Loosdrecht, D.~I. Khomskii, J.~van~den Brink, and
  M.~Gr{\"u}ninger.
\newblock Resonant inelastic x-ray incarnation of young's double-slit
  experiment.
\newblock {\em Science Advances}, 5(1):eaav4020, 2019.

\bibitem{Magnaterra2023}
M.~Magnaterra, M.~Moretti~Sala, G.~Monaco, P.~Becker, M.~Hermanns,
  P.~Warzanowski, T.~Lorenz, D.~I. Khomskii, P.~H.~M. van Loosdrecht,
  J.~van~den Brink, and M.~Gr\"uninger.
\newblock Rixs interferometry and the role of disorder in the quantum magnet
  ${\mathrm{ba}}_{3}{\mathrm{ti}}_{3\ensuremath{-}x}{\mathrm{ir}}_{x}{\mathrm{o}}_{9}$.
\newblock {\em Phys. Rev. Res.}, 5:013167, Mar 2023.

\bibitem{sync2}
P.R. Willmott.
\newblock {\em An Introduction to Synchrotron Radiation: Techniques and
  Applications}.
\newblock Wiley, Hoboken, 2019.

\bibitem{McN10:814}
Brian W.~J. McNeil and Neil~R. Thompson.
\newblock X-ray free-electron lasers.
\newblock {\em Nature Photonics}, 4(12):814--821, December 2010.

\bibitem{Pel16:15006}
C.~Pellegrini, A.~Marinelli, and S.~Reiche.
\newblock The physics of x-ray free-electron lasers.
\newblock {\em Rev. Mod. Phys.}, 88:015006, Mar 2016.

\bibitem{Zha17:}
Zhentang Zhao, Dong Wang, Qiang Gu, Lixin Yin, Ming Gu, Yongbin Leng, and
  Bo~Liu.
\newblock Status of the sxfel facility.
\newblock {\em Applied Sciences}, 7(6), 2017.

\bibitem{Sed17:115901}
E~A Seddon, J~A Clarke, D~J Dunning, C~Masciovecchio, C~J Milne, F~Parmigiani,
  D~Rugg, J~C~H Spence, N~R Thompson, K~Ueda, S~M Vinko, J~S Wark, and W~Wurth.
\newblock Short-wavelength free-electron laser sources and science: a review.
\newblock {\em Reports on Progress in Physics}, 80(11):115901, oct 2017.

\bibitem{Beye2013a}
M.~Beye, Ph. Wernet, C.~Sch{\"u}{\ss}ler-Langeheine, and A.~F{\"o}hlisch.
\newblock Time resolved resonant inelastic x-ray scattering: A supreme tool to
  understand dynamics in solids and molecules.
\newblock {\em Journal of Electron Spectroscopy and Related Phenomena},
  188:172--182, 2013.
\newblock Progress in Resonant Inelastic X-Ray Scattering.

\bibitem{Lu2020}
H.~Lu, A.~Gauthier, M.~Hepting, A.~S. Tremsin, A.~H. Reid, P.~S. Kirchmann,
  Z.~X. Shen, T.~P. Devereaux, Y.~C. Shao, X.~Feng, G.~Coslovich, Z.~Hussain,
  G.~L. Dakovski, Y.~D. Chuang, and W.~S. Lee.
\newblock Time-resolved rixs experiment with pulse-by-pulse parallel readout
  data collection using x-ray free electron laser.
\newblock {\em Scientific Reports}, 10(1):22226, 2020.

\bibitem{Gel2021}
Faris Gel'mukhanov, Michael Odelius, Sergey~P. Polyutov, Alexander F\"ohlisch,
  and Victor Kimberg.
\newblock Dynamics of resonant x-ray and auger scattering.
\newblock {\em Rev. Mod. Phys.}, 93:035001, Jul 2021.

\bibitem{Mitrano2020}
Matteo Mitrano and Yao Wang.
\newblock Probing light-driven quantum materials with ultrafast resonant
  inelastic x-ray scattering.
\newblock {\em Communications Physics}, 3(1):184, 2020.

\bibitem{Wernet2015}
Ph. Wernet, K.~Kunnus, I.~Josefsson, I.~Rajkovic, W.~Quevedo, M.~Beye,
  S.~Schreck, S.~Gr{\"u}bel, M.~Scholz, D.~Nordlund, W.~Zhang, R.~W. Hartsock,
  W.~F. Schlotter, J.~J. Turner, B.~Kennedy, F.~Hennies, F.~M.~F. de~Groot,
  K.~J. Gaffney, S.~Techert, M.~Odelius, and A.~F{\"o}hlisch.
\newblock Orbital-specific mapping of the ligand exchange dynamics of fe(co)5
  in solution.
\newblock {\em Nature}, 520(7545):78--81, 2015.

\bibitem{Kunnus2016}
Kristjan Kunnus, Ida Josefsson, Ivan Rajkovic, Simon Schreck, Wilson Quevedo,
  Martin Beye, Sebastian Gr{\"u}bel, Mirko Scholz, Dennis Nordlund, Wenkai
  Zhang, Robert~W Hartsock, Kelly~J Gaffney, William~F Schlotter, Joshua~J
  Turner, Brian Kennedy, Franz Hennies, Simone Techert, Philippe Wernet,
  Michael Odelius, and Alexander F{\"o}hlisch.
\newblock Anti-stokes resonant x-ray raman scattering for atom specific and
  excited state selective dynamics.
\newblock {\em New Journal of Physics}, 18(10):103011, oct 2016.

\bibitem{Kunnus2016b}
K.~Kunnus, I.~Josefsson, I.~Rajkovic, S.~Schreck, W.~Quevedo, M.~Beye,
  C.~Weniger, S.~Grübel, M.~Scholz, D.~Nordlund, W.~Zhang, R.~W. Hartsock,
  K.~J. Gaffney, W.~F. Schlotter, J.~J. Turner, B.~Kennedy, F.~Hennies,
  F.~M.~F. de~Groot, S.~Techert, M.~Odelius, Ph. Wernet, and A.~Föhlisch.
\newblock {Identification of the dominant photochemical pathways and
  mechanistic insights to the ultrafast ligand exchange of Fe(CO)5 to
  Fe(CO)4EtOH}.
\newblock {\em Structural Dynamics}, 3(4):043204, 02 2016.

\bibitem{Jay2018}
Raphael~M. Jay, Jesper Norell, Sebastian Eckert, Markus Hantschmann, Martin
  Beye, Brian Kennedy, Wilson Quevedo, William~F. Schlotter, Georgi~L.
  Dakovski, Michael~P. Minitti, Matthias~C. Hoffmann, Ankush Mitra, Stefan~P.
  Moeller, Dennis Nordlund, Wenkai Zhang, Huiyang~W. Liang, Kristjan Kunnus,
  Katharina Kubi{\v c}ek, Simone~A. Techert, Marcus Lundberg, Philippe Wernet,
  Kelly Gaffney, Michael Odelius, and Alexander F{\"o}hlisch.
\newblock Disentangling transient charge density and metal--ligand covalency in
  photoexcited ferricyanide with femtosecond resonant inelastic soft x-ray
  scattering.
\newblock {\em The Journal of Physical Chemistry Letters}, 9(12):3538--3543, 06
  2018.

\bibitem{Norell2018}
Jesper Norell, Raphael~M. Jay, Markus Hantschmann, Sebastian Eckert, Meiyuan
  Guo, Kelly~J. Gaffney, Philippe Wernet, Marcus Lundberg, Alexander
  F{\"o}hlisch, and Michael Odelius.
\newblock Fingerprints of electronic{,} spin and structural dynamics from
  resonant inelastic soft x-ray scattering in transient photo-chemical species.
\newblock {\em Phys. Chem. Chem. Phys.}, 20:7243--7253, 2018.

\bibitem{Jay2021}
Raphael~M. Jay, Sebastian Eckert, Benjamin~E. Van~Kuiken, Miguel Ochmann,
  Markus Hantschmann, Amy~A. Cordones, Hana Cho, Kiryong Hong, Rory Ma,
  Jae~Hyuk Lee, Georgi~L. Dakovski, Joshua~J. Turner, Michael~P. Minitti,
  Wilson Quevedo, Annette Pietzsch, Martin Beye, Tae~Kyu Kim, Robert~W.
  Schoenlein, Philippe Wernet, Alexander F{\"o}hlisch, and Nils Huse.
\newblock Following metal-to-ligand charge-transfer dynamics with ligand and
  spin specificity using femtosecond resonant inelastic x-ray scattering at the
  nitrogen k-edge.
\newblock {\em The Journal of Physical Chemistry Letters}, 12(28):6676--6683,
  2021.
\newblock PMID: 34260255.

\bibitem{Eckert2017}
Sebastian Eckert, Jesper Norell, Piter~S. Miedema, Martin Beye, Mattis Fondell,
  Wilson Quevedo, Brian Kennedy, Markus Hantschmann, Annette Pietzsch,
  Benjamin~E. Van~Kuiken, Matthew Ross, Michael~P. Minitti, Stefan~P. Moeller,
  William~F. Schlotter, Munira Khalil, Michael Odelius, and Alexander
  Föhlisch.
\newblock Ultrafast independent n-h and n-c bond deformation investigated with
  resonant inelastic x-ray scattering.
\newblock {\em Angewandte Chemie International Edition}, 56(22):6088--6092,
  2017.

\bibitem{Norman2018}
Patrick Norman and Andreas Dreuw.
\newblock Simulating x-ray spectroscopies and calculating core-excited states
  of molecules.
\newblock {\em Chemical Reviews}, 118(15):7208--7248, 08 2018.

\bibitem{Kramers1925}
H.~A. Kramers and W.~Heisenberg.
\newblock Über die streuung von strahlung durch atome.
\newblock {\em Zeitschrift für Physik}, 31(1):681--708, feb 1925.

\bibitem{Dirac1927}
P.A.M. Dirac and R.H. Fowler.
\newblock The quantum theory of dispersion.
\newblock {\em Proceedings of the Royal Society of London. Series A, Containing
  Papers of a Mathematical and Physical Character}, 114(769):710--728, may
  1927.

\bibitem{Lee1979}
Soo‐Y. Lee and E.~J. Heller.
\newblock {Time‐dependent theory of Raman scattering}.
\newblock {\em The Journal of Chemical Physics}, 71(12):4777--4788, 07 1979.

\bibitem{Tannor1982}
David~J. Tannor and Eric~J. Heller.
\newblock Polyatomic raman scattering for general harmonic potentials.
\newblock {\em The Journal of Chemical Physics}, 77(1):202--218, jul 1982.

\bibitem{Banerjee2024}
Ambar Banerjee, Raphael~M. Jay, Torsten Leitner, Ru-Pan Wang, Jessica Harich,
  Robert Stefanuik, Michael~R. Coates, Emma~V. Beale, Victoria Kabanova,
  Abdullah Kahraman, Anna Wach, Dmitry Ozerov, Christopher Arrell, Christopher
  Milne, Philip J.~M. Johnson, Claudio Cirelli, Camila Bacellar, Nils Huse,
  Michael Odelius, and Philippe Wernet.
\newblock Accessing metal-specific orbital interactions in c--h activation with
  resonant inelastic x-ray scattering.
\newblock {\em Chem. Sci.}, 15:2398--2409, 2024.

\bibitem{Wang03}
Haobin Wang and Michael Thoss.
\newblock {Multilayer formulation of the multiconfiguration time-dependent
  Hartree theory}.
\newblock {\em The Journal of Chemical Physics}, 119(3):1289--1299, 07 2003.

\bibitem{Manthe08}
Uwe Manthe.
\newblock {A multilayer multiconfigurational time-dependent Hartree approach
  for quantum dynamics on general potential energy surfaces}.
\newblock {\em The Journal of Chemical Physics}, 128(16):164116, 04 2008.

\bibitem{Vendrell11}
Oriol Vendrell and Hans-Dieter Meyer.
\newblock {Multilayer multiconfiguration time-dependent Hartree method:
  Implementation and applications to a Henon–Heiles Hamiltonian and to
  pyrazine}.
\newblock {\em The Journal of Chemical Physics}, 134(4):044135, 01 2011.

\bibitem{Beck00}
M.H. Beck, A.~J{\"a}ckle, G.A. Worth, and H.-D. Meyer.
\newblock The multiconfiguration time-dependent hartree (mctdh) method: a
  highly efficient algorithm for propagating wavepackets.
\newblock {\em Physics Reports}, 324(1):1--105, 2000.

\bibitem{Mey09:}
Graham A.~Worth Hans-Dieter~Meyer, Fabien~Gatti, editor.
\newblock {\em Multidimensional Quantum Dynamics: MCTDH Theory and
  Applications}.
\newblock WILEY-VCH Verlag GmbH \& Co. KGaA, 2009.

\bibitem{Dirac1930}
P.~A.~M. Dirac.
\newblock Note on exchange phenomena in the thomas atom.
\newblock {\em Mathematical Proceedings of the Cambridge Philosophical
  Society}, 26(3):376–385, 1930.

\bibitem{Fre34:}
J.~Frenkel.
\newblock {\em Wave mechanics; advanced general theory}.
\newblock The international series of monographs on physics. The Clarendon
  press, Oxford, 1934.

\bibitem{Koeppel84}
H.~Köppel, W.~Domcke, and L.~S. Cederbaum.
\newblock {\em Multimode Molecular Dynamics Beyond the Born-Oppenheimer
  Approximation}, pages 59--246.
\newblock John Wiley $\&$ Sons, Ltd, 1984.

\bibitem{Worth04}
Graham~A. Worth and Lorenz~S. Cederbaum.
\newblock Beyond born-oppenheimer: Molecular dynamics through a conical
  intersection.
\newblock {\em Annual Review of Physical Chemistry}, 55(1):127--158, 2004.
\newblock PMID: 15117250.

\bibitem{Cederbaum1977}
L.S. Cederbaum, W.~Domcke, H.~K{\"o}ppel, and W.~{Von Niessen}.
\newblock Strong vibronic coupling effects in ionization spectra: The ``mystery
  band'' of butatriene.
\newblock {\em Chemical Physics}, 26(2):169--177, 1977.

\bibitem{Domcke1977}
W.~Domcke and L.S. Cederbaum.
\newblock Vibronic coupling and symmetry breaking in core electron ionization.
\newblock {\em Chemical Physics}, 25(2):189--196, 1977.

\bibitem{Lee1989xas}
Soo-Y. Lee, W.Thomas Pollard, and Richard~A. Mathies.
\newblock Quantum theory for transition state absorption.
\newblock {\em Chemical Physics Letters}, 160(5):531--537, 1989.

\bibitem{Freibert2024}
Antonia Freibert, David Mendive-Tapia, Nils Huse, and Oriol Vendrell.
\newblock Time-dependent resonant inelastic x-ray scattering of pyrazine at the
  nitrogen k-edge: A quantum dynamics approach.
\newblock {\em Journal of Chemical Theory and Computation}, 02 2024.

\bibitem{Sala14}
Matthieu Sala, Benjamin Lasorne, Fabien Gatti, and St{\'e}phane Gu{\'e}rin.
\newblock The role of the low-lying dark n$\pi$* states in the photophysics of
  pyrazine: a quantum dynamics study.
\newblock {\em Phys. Chem. Chem. Phys.}, 16:15957--15967, 2014.

\bibitem{Vid19:3117}
Marta Vidal, Xintian Feng, Evgeny Epifanovsky, Anna~I. Krylov, and Sonia
  Coriani.
\newblock New and efficient equation-of-motion coupled-cluster framework for
  core-excited and core-ionized states.
\newblock {\em J. Chem. Theory Comput.}, 15(5):3117--3133, May 2019.

\bibitem{Dunning1989}
Thom~H. Dunning.
\newblock Gaussian basis sets for use in correlated molecular calculations. i.
  the atoms boron through neon and hydrogen.
\newblock {\em The Journal of Chemical Physics}, 90(2):1007--1023, 1989.

\bibitem{qchem5}
Evgeny Epifanovsky and et~al.
\newblock Software for the frontiers of quantum chemistry: An overview of
  developments in the q-chem 5 package.
\newblock {\em J. Chem. Phys.}, 155 (8), 2021.

\bibitem{mctdh}
G.~A. Worth, M.~H. Beck, A.~J\"ackle, H.–D. Meyer, F.~Otto, M.~Brill, and
  O.~Vendrell.
\newblock The mctdh package, version 8.6, 2020.

\bibitem{Pri99:141}
K.C. Prince, M.~Vondráček, J.~Karvonen, M.~Coreno, R.~Camilloni, L.~Avaldi,
  and M.~{de Simone}.
\newblock A critical comparison of selected 1s and 2p core hole widths.
\newblock {\em Journal of Electron Spectroscopy and Related Phenomena},
  101-103:141--147, 1999.

\bibitem{Freibert2021}
Antonia Freibert, David Mendive-Tapia, Nils Huse, and Oriol Vendrell.
\newblock Femtosecond x-ray absorption spectroscopy of pyrazine at the nitrogen
  k-edge: on the validity of the lorentzian limit.
\newblock {\em Journal of Physics B: Atomic, Molecular and Optical Physics},
  54(24):244003, dec 2021.

\bibitem{LCLSII}
P.~Abbamonte, Frank Abild-Pedersen, P.~Adams, M.~Ahmed, F.~Albert, R.~Alonso
  Mori, P.~Anfinrud, A.~Aquila, M.~Armstrong, J.~Arthur, J.~Bargar, A.~Barty,
  U.~Bergmann, N.~Berrah, G.~Blaj, H.~Bluhm, C.~Bolme, C.~Bostedt, S.~Boutet,
  G.~Brown, P.~Bucksbaum, M.~Cargnello, G.~Carini, A.~Cavalleri, V.~Cherezov,
  W.~Chiu, Y.~Chuang, D.~Cocco, R.~Coffee, G.~Collins, A.~Cordones-Hahn,
  J.~Cryan, G.~Dakovski, M.~Dantus, H.~Demirci, P.~Denes, T.~Devereaux,
  Y.~Ding, S.~Doniach, R.~Dorner, M.~Dunne, H.~Durr, T.~Egami, D.~Eisenberg,
  P.~Emma, C.~Fadley, R.~Falcone, Y.~Feng, P.~Fischer, F.~Fiuza, L.~Fletcher,
  L.~Foucar, M.~Frank, J.~Fraser, H.~Frei, D.~Fritz, P.~Fromme, A.~Fry,
  M.~Fuchs, P.~Fuoss, K.~Gaffney, E.~Gamboa, O.~Gessner, S.~Ghimire,
  A.~Gleason, S.~Glenzer, T.~Gorkhover, A.~Gray, M.~Guehr, J.~Guo, J.~Hajdu,
  S.~Hansen, P.~Hart, M.~Hashimoto, J.~Hastings, D.~Haxton, P.~Heimann,
  T.~Heinz, A.~Hexemer, J.~Hill, F.~Himpsel, P.~Ho, B.~Hogue, Z.~Huang,
  M.~Hunter, G.~Hura, N.~Huse, Z.~Hussain, M.~Ilchen, C.~Jacobsen, C.~Kenney,
  J.~Kern, S.~Kevan, J.~Kim, H.~Kim, P.~Kirchmann, R.~Kirian, S.~Kivelson,
  C.~Kliewer, J.~Koralek, G.~Kovacsova, A.~Lanzara, J.~LaRue, H.~Lee, J.~Lee,
  W.~Lee, Y.~Lee, I.~Lindau, A.~Lindenberg, Z.~Liu, D.~Lu, U.~Lundstrom,
  A.~MacDowell, W.~Mao, J.~Marangos, G.~Marcus, T.~Martinez, W.~McCurdy,
  G.~McDermott, C.~McGuffey, M.~Minitti, S.~Miyabe, S.~Moeller, R.~Moore,
  S.~Mukamel, K.~Nass, A.~Natan, K.~Nelson, S.~Nemsak, D.~Neumark, R.~Neutze,
  A.~Nilsson, D.~Nordlund, J.~Norskov, S.~Nozawa, H.~Ogasawara, H.~Ohldag,
  A.~Orville, D.~Osborn, T.~Osipov, A.~Ourmazd, D.~Parkinson, C.~Pellegrini,
  G.~Phillips, T.~Rasing, T.~Raubenheimer, T.~Recigno, A.~Reid, D.~Reis,
  A.~Robert, J.~Robinson, D.~Rolles, J.~Rost, S.~Roy, A.~Rudenko, T.~Russell,
  R.~Sandberg, A.~Sandhu, N.~Sauter, I.~Schlichting, R.~Schlogl, W.~Schlotter,
  M.~Schmidt, J.~Schneider, R.~Schoenlein, M.~Schoeffler, A.~Scholl, Z.~Shen,
  O.~Shpyrko, T.~Silva, S.~Sinha, D.~Slaughter, J.~Sobota, D.~Sokaras,
  K.~Sokolowski-Tinten, S.~Southworth, J.~Spence, C.~Stan, J.~Stohr, R.~Stroud,
  V.~Sundstrom, C.~Taatjes, A.~Thomas, M.~Trigo, Y.~Tsui, J.~Turner, A.~van
  Buuren, S.~Vinko, S.~Wakatsuki, J.~Wark, P.~Weber, T.~Weber, M.~Wei,
  T.~Weiss, P.~Wernet, W.~White, P.~Willmott, K.~Wilson, W.~Wurth,
  V.~Yachandra, J.~Yano, D.~Yarotski, L.~Young, Y.~Zhu, D.~Zhu, and P.~Zwart.
\newblock New science opportunities enabled by lcls-ii x-ray lasers.
\newblock 6 2015.

\end{thebibliography}

\end{document}